# Predicting frictional ageing from bulk relaxation measurements

Kasra Farain and Daniel Bonn

Van der Waals–Zeeman Institute, Institute of Physics, University of Amsterdam, Science Park 904, 1098 XH Amsterdam, The Netherlands.

**Abstract:** The coefficient of static friction between solids generally depends on the time they have remained in static contact before the measurement. Such frictional aging is at the origin of the difference between static and dynamic friction coefficients, but has remained difficult to understand. It is usually attributed to a slow increase in the area of atomic contact as the interface changes under pressure. This is however very difficult to quantify as surfaces have roughness at all length scales, and friction is not always proportional to the contact area. Here, we show that plastic flow of surface irregularities within a polymer-on-glass frictional interface exhibits identical relaxation dynamics as that of the bulk, allowing to predict the rate of frictional aging.

The contact between common solids involves time-dependent localized surface deformations. When two rough surfaces are pressed together, the mechanical interaction between them occurs through a force-bearing ensemble of island-like contact spots (*1-7*). Depending on the surface roughness and material characteristics, the local stress on the contact area can easily exceed the yield stress of the material. As a result, some parts of the contact area will exhibit plastic flow while other less-strained parts will contribute to supporting the total normal load elastically (*6-8*). This highly complex elastoplastic problem is at the heart of a fundamental understanding of many important surface and interface phenomena, from friction and adhesion to sealing and interfacial stiffness.

For friction, Amontons' law states that the frictional force between two bodies is proportional to the normal load acting between them and is independent of the macroscopic area of contact. These are usually understood as a consequence of the surface roughness: the area of atomic contact between the two bodies *A* is only a small fraction of the macroscopic contact area, and can be proportional to the normal load *F* in some simple models. Assuming that the overall shear strength of an interface is constant during sliding, one recovers Amontons' law (*1, 6, 9*). For over a century, various experimental techniques from electrical conductance measurements to optical and recently fluorescence observations have investigated the relation between *A* and *F* under a broad range of conditions (*6, 9-11*). However, a full analytical or computational understanding of this problem, even when reduced to an equilibrium problem, is still lacking. Early work focused on elastic deformations of independent surface asperities (*12-14*), but it has recently become clear that this severely underestimates the real area of contact, because plastic flow is neglected (*7*). More recently, a contact mechanics theory was developed by Persson that can take into account plastic deformations by supposing that wherever the yield criterion (e.g., the von Mises yield condition) of the softer solid is satisfied, (instant) plastic changes occur (*5, 15, 16*). Finite-element calculations also use such a fixed yield criterion (*7, 17-20*). This however cannot account for experimentally observed slow aging dynamics of static frictional interfaces that leads to the difference between static and dynamic friction. In the simplest scenario and under a constant load, the contact area and frictional strength grow simultaneously with the logarithm of time (*6, 10*). Consequently, frictional aging is usually considered as being solely the result of contact area creep (*10, 21*). However, recent work shows that in reality friction force and area may even evolve in different directions (*6, 22*). Further hampering our understanding of frictional aging is due to indications that the glass transition temperature (*23-27*) and, therefore, the molecular mobility may be different close to surface or interface from that in the bulk. For example, the structural relaxation rate of poly(methyl methacrylate), measured by fluorescence probe molecules, can decrease by a factor of 2 at a free surface and by a factor of 15 at a silica interface (*28*). Settling these issues is all the more important as the slow evolution of frictional interfaces constitutes the basis of transient sliding dynamics at low velocities (e.g. rate and state friction transients), emerging frictional instabilities, and slip-stick phenomena (*1, 4, 29*).

In the current work, we investigate the evolution of friction due to plastic flow and stress relaxation of surface irregularities in glassy polymers in contact with a rigid surface. We first study the stress relaxation of polymer glasses after macroscale bulk and/or microscale surface plastic deformations. We find a simple and generic expression for the stress relaxation dynamics without any significant signature of the surface structure. We then show that the same time-dependence also emerges from friction measurements, indicating that the glassy relaxation determines the frictional aging.

Our setup for stress relaxation experiments is schematically shown in Fig. 1A. A polymer sphere is held gently between a steel and a glass plate. The polymer materials used here are polypropylene and polytetrafluoroethylene that have existed in their metastable equilibrium state for a long time before the experiments. The top steel plate is attached to a rheometer to measure the pressing force. The rheometer has a large internal spring constant so that the gap doesn't change after the stress relaxation in polymer spheres. For the same reason, the bottom glass slide is attached to a strong custom-made frame. We decrease the gap (at a speed of roughly 1 mm/s) to squeeze the polymer sphere while the rheometer measures the normal force $F$ and the microscope takes images (see Methods).

Two typical examples of the relaxation of the force that the sphere exerts on the plates as a function of time $t$ are shown in Fig. 1B. Under a small load of 0.12 N, only surface plastic deformations are possible. On the other hand, under 53.05 N load the bulk of the sphere deforms non-reversibly (see Supp. Mat.). However, in both cases, after a brief onset ($t < 0.5$ s) the force relaxation is logarithmic over many decades in time:

$$F(t) = B - C \ln\left(\frac{t}{1\,s}\right), \tag{1}$$

where $C$ and $B$ are constants. Previously, other molecular and/or athermal disordered systems including creased sheets (*30, 31*), crumpled papers (*32, 33*), and foams (*33*) were also reported to display such logarithmic mechanical response to deformation. Moreover, enthalpy and specific volume measurements during aging of polymers show a similar logarithmic behavior (*34*). In the case of creased polymeric (Mylar and paper) sheets, which contain lines of localized plastic deformations, analysis of the relationship between $C$ and $B$ revealed that the logarithmic process is governed by a constant that doesn't change with the sheet thickness or the applied load (*31*). Similarly, for our polymer spheres, $B$ is proportional to $C$ for the whole range of the squeezing forces, from 0.1 N to above 50 N (Fig. 2A). As a result, we can write $B = C \ln(\frac{\tau}{1\,s})$, where $\ln(\frac{\tau}{1\,s})$ is the slope of the plot, irrespective of whether bulk or surface plastic deformations occur. Putting this equation into Eq. 1, $F$ can be expressed as $F = -C \ln\left(\frac{t}{\tau}\right)$. Then, in Fig. 2B, we plot $C$ as a function of the initially applied force $F_0$, which again yields a simple linear proportionality, $C = nF_0$, so that:

$$F(t) = F_0 \ln\left(\frac{t}{\tau}\right)^{-n}. \tag{2}$$

This expression is valid over the full range of forces, even though the corresponding contact areas are widely different (Fig. 2C). For example, the contact area corresponding to $F_0 = 0.12$ N consists of small islands determined by the surface roughness of the sphere while the contact image corresponding to $F_0 = 53.05$ N displays an almost complete Hertzian-type contact circle.

To determine when the transition from elastic to plastic deformations occurs in these experiments, we plot the total deformation $d$ of the sphere against the initial compressive force $F_0$ exerted on it (Fig. 3). For small forces, we find $F_0 \propto d^{3/2}$ which corresponds to Hertz' contact theory for bulk elastic deformation. For larger forces, the plot shows a transition to a linear behavior associated with plastic deformation. The intercept of fits to the elastic and plastic regimes can be considered as the point where the bulk of the sphere yields to flow (*8*).

Next, we focus on the small compressive force regime ($F_0 < 10$ N), where plastic deformations are limited to the sphere's surface asperities. The surprise is that any configuration of the real contact area (determined by the details of the surface irregularities) behaves according to Eq. 2. This signifies that Eq. 2 is also valid locally. Hence, in a frictional contact with a distribution of contact pressures (*35*), all microscopic elastic forces acting within the interface will dissipate at the same rate everywhere.

We now investigate whether the frictional strength, as a collective macroscopic observable of the evolution of a static interface, also evolves at the same rate as the relaxation of the local normal stresses. Our apparatus for friction measurements is shown in Fig. 4A. We use a rheometer with a custom-made tool to rotate a cylindrical tube with three polymer spheres glued to its underside, while it is measuring the applied torque. Figure 4B shows typical examples of friction as a function of sliding distance for polypropylene spheres on a glass substrate with different waiting times $t_w$ before the measurement. As generally expected, the static or peak friction $F_s$ grows with waiting time $t_w$. Fig. 4C plots $F_s$ normalized by the dynamic or steady-state friction $F_d$ as a function of aging time $t_w$ for multiple experiments. These data are in close agreement with the structural relaxation rate obtained from squeezing experiments (solid line), indicating that the frictional aging is given by the plastic surface deformations. This effect is nearly independent of the details of the surface topography, the imposed sliding velocity, and the underlying surface interactions. For example, very similar frictional aging is observed for polypropylene on both glass and silicon substrates.

Our results for polypropylene spheres (and repeated for polytetrafluoroethylene spheres, see Supp. Mat.) not only demonstrate that the aging of frictional interfaces of polymer glasses originates from structural relaxation of the glassy state, but also provide a quantitative description of these relaxation dynamics. We obtain a structural relaxation constant from bulk relaxation measurements that directly predicts the evolution of the static friction in time. The observation that the microscopic compressive forces within plastically deformed areas in contact interfaces undergo a generic material-dependent relaxation could also provide a basis for a theoretical understanding of the time-dependence of macroscopic adhesion between solids (*2, 5, 36*). Especially, the adhesion hysteresis which has already been qualitatively attributed to viscoelastic dissipation can be addressed through our results.

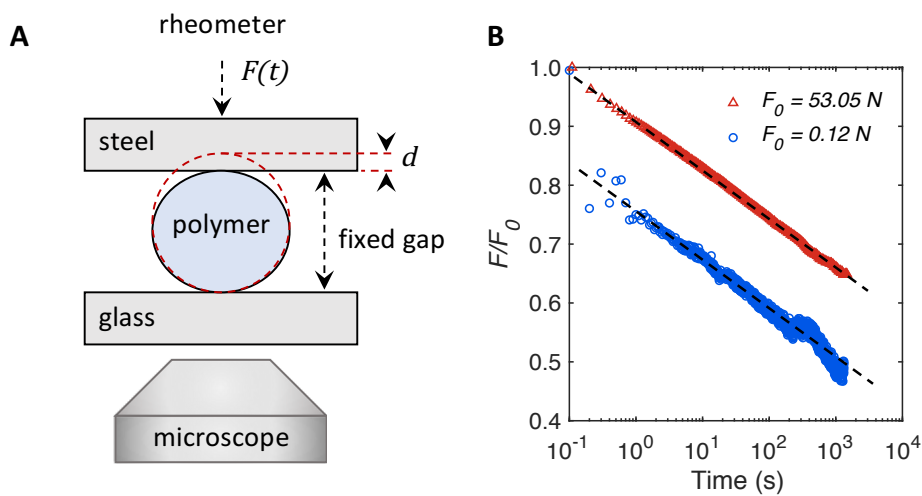

Fig. 1. (**A**) Schematic of the setup for deformation experiments. A polypropylene or polytetrafluoroethylene sphere is squeezed between a steel plate connected to a rheometer and a glass slide installed on a microscope via a very stiff frame. (**B**) Relaxation of the applied force $F$ needed to maintain a constant strain as a function of time for two values of the initial force $F_0$; 0.12 N (blue circles) and 53.05 N (red triangles). Dashed lines indicate logarithmic relaxation, Eq. 1.

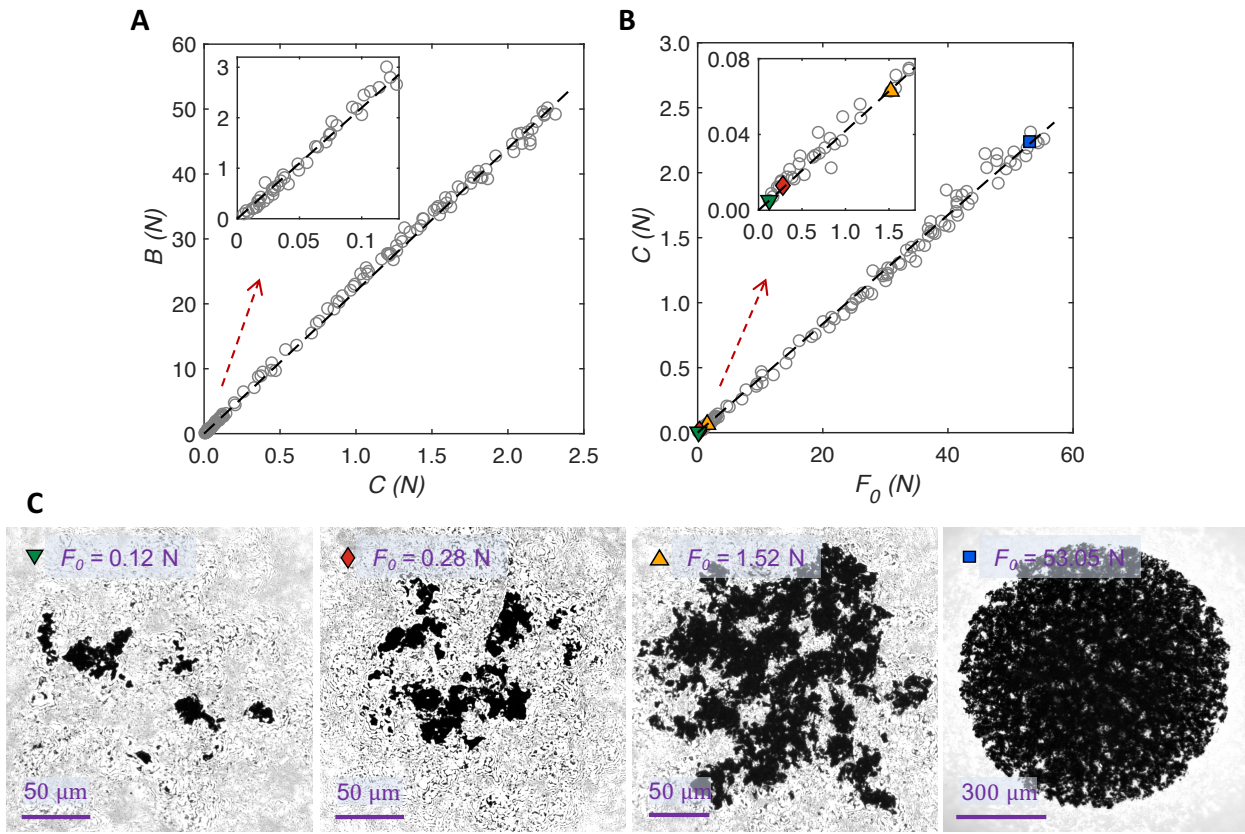

Fig. 2. Relation between constants $C$ and $B$ (**A**) and constant $C$ and $F_0$ (**B**) as obtained from fitting stress relaxation characteristics to Eq. 1 for different values of initially applied normal force $F_0$. Dashed lines are linear fits. (**C**) Microscopy images of the contact area for four values of $F_0$. Colored symbols indicate which data points in (**B**) correspond to which images in (**C**).

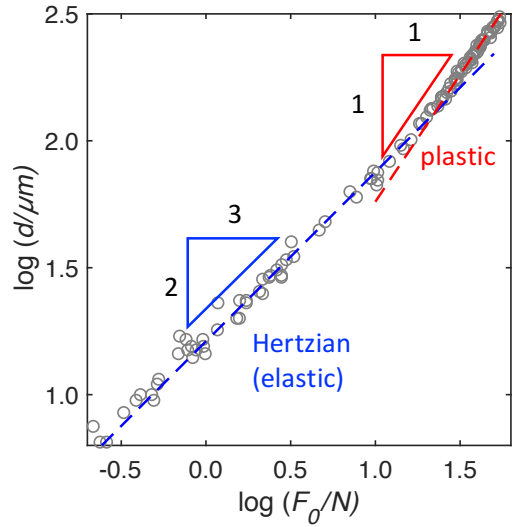

Fig. 3. Deformation $d$ of a polypropylene sphere subjected to an initial compressive force $F_0$. As $d$ is (much) larger than the surface roughness of the sphere (< 1 µm), it represents a bulk parameter. At small forces ($F_0 < 10$ N), $d$ increases with $F_0$ to the power of 2/3 (blue dashed line), corresponding to Hertz' law for elastic deformation. At larger forces this changes into a linear increase (red dashed line), corresponding to plastic deformation.

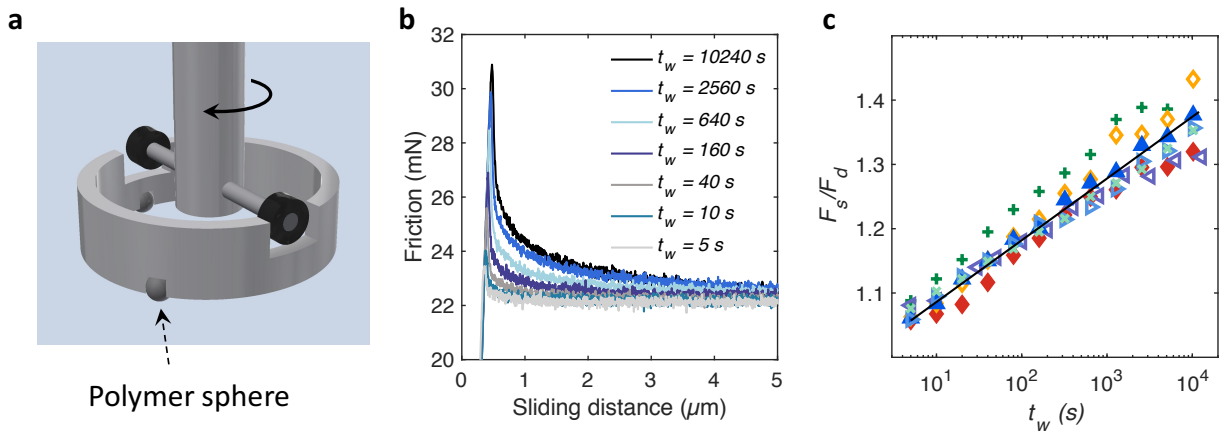

Fig. 4. (**A**) Schematic of the apparatus for the friction measurements. A rheometer is used to rotate a hollow tube around its symmetry axis. Three polymer spheres are attached to the underside of the tube equidistantly which with a bottom substrate form the frictional interfaces. (**B**) Friction as a function of sliding distance for polypropylene spheres on glass under the combined weight of the tube and spheres (42 mN) after different waiting times $t_w$. The polymer spheres are not changed between these measurements. (**C**) The ratio of static friction ($F_s$, the peak friction value) to dynamic friction ($F_d$, mean friction in steady-state) for multiple aging experiments as in (**A**): solid triangles are from curves in (**A**), open right triangles correspond to a subsequent repeat of the experiment with the same spheres, open left triangles are with new polypropylene spheres, × signs are with a new tube of 87 mN weight, and red diamonds

correspond to an experiment on a silicon wafer. In all the above experiments, the imposed sliding velocity is 86 nm/s. + signs and open diamonds are repeats of the last two experiments, respectively, with the sliding velocity of 258 nm/s. The solid line corresponds to equation $F_s(t) = F_\tau - F_d \ln\left(\frac{t}{\tau}\right)^{-n}$ with $n = 0.0417$ determined from Fig. 2B (experimental value: $n = 0.0417 \pm 0.0002$). The free parameter $F_\tau$ is the static friction at time $\tau$ (for polypropylene $\tau = 135 \pm 8$ years, obtained from Fig. 2A).

**References**

1. B. N. J. Persson, Sliding Friction Physical Principles and Applications. *(Springer, New York, 2000)*.
2. A. G. Peressadko, N. Hosoda, B. N. J. Persson, Influence of Surface Roughness on Adhesion between Elastic Bodies. *Physical Review Letters* **95**, 124301 (2005).
3. B. N. J. Persson, O. Albohr, U. Tartaglino, A. I. Volokitin, E. Tosatti, On the nature of surface roughness with application to contact mechanics, sealing, rubber friction and adhesion. *Journal of Physics: Condensed Matter* **17**, R1-R62 (2004).
4. T. Baumberger, C. Caroli, Solid friction from stick–slip down to pinning and aging. *Advances in Physics* **55**, 279-348 (2006).
5. B. N. J. Persson, Elastoplastic Contact between Randomly Rough Surfaces. *Physical Review Letters* **87**, 116101 (2001).
6. B. Weber, T. Suhina, A. M. Brouwer, D. Bonn, Frictional weakening of slip interfaces. *Science Advances* **5**, eaav7603 (2019).
7. B. Weber *et al.*, Molecular probes reveal deviations from Amontons' law in multi-asperity frictional contacts. *Nature Communications* **9**, 888 (2018).
8. A. Tiwari, a. wang, M. Müser, B. Persson, Contact Mechanics for Solids with Randomly Rough Surfaces and Plasticity. *Lubricants* **7**, (2019).
9. F. P. Bowden, D. Tabor, G. I. Taylor, The area of contact between stationary and moving surfaces. *Proceedings of the Royal Society of London. Series A. Mathematical and Physical Sciences* **169**, 391-413 (1939).
10. J. H. Dieterich, B. D. Kilgore, Direct observation of frictional contacts: New insights for state-dependent properties. *Pure Appl. Geophys.* **143**, 283-302 (1994).
11. T. Suhina *et al.*, Fluorescence Microscopy Visualization of Contacts Between Objects. *Angewandte Chemie International Edition* **54**, 3688-3691 (2015).
12. J. A. Greenwood, J. B. P. Williamson, Contact of Nominally Flat Surfaces. *Proceedings of the Royal Society of London. Series A, Mathematical and Physical Sciences* **295**, 300-319 (1966).
13. A. W. Bush, R. D. Gibson, T. R. Thomas, The elastic contact of a rough surface. *Wear* **35**, 87-111 (1975).
14. G. Carbone, F. Bottiglione, Asperity contact theories: Do they predict linearity between contact area and load? *Journal of the Mechanics and Physics of Solids* **56**, 2555-2572 (2008).

15. B. N. J. Persson, Theory of rubber friction and contact mechanics. *The Journal of Chemical Physics* **115**, 3840-3861 (2001).
16. W. B. Dapp, N. Prodanov, M. H. Müser, Systematic analysis of Persson's contact mechanics theory of randomly rough elastic surfaces. *Journal of Physics: Condensed Matter* **26**, 355002 (2014).
17. B. Luan, M. O. Robbins, The breakdown of continuum models for mechanical contacts. *Nature* **435**, 929-932 (2005).
18. A. Almqvist, F. Sahlin, R. Larsson, S. Glavatskih, On the dry elasto-plastic contact of nominally flat surfaces. *Tribology International* **40**, 574-579 (2007).
19. L. Pei, S. Hyun, J. F. Molinari, M. O. Robbins, Finite element modeling of elasto-plastic contact between rough surfaces. *Journal of the Mechanics and Physics of Solids* **53**, 2385-2409 (2005).
20. M. H. Müser *et al.*, Meeting the Contact-Mechanics Challenge. *Tribology Letters* **65**, 118 (2017).
21. S. Dillavou, S. M. Rubinstein, Shear Controls Frictional Aging by Erasing Memory. *Physical Review Letters* **124**, 085502 (2020).
22. S. Dillavou, S. M. Rubinstein, Nonmonotonic Aging and Memory in a Frictional Interface. *Physical Review Letters* **120**, 224101 (2018).
23. C. J. Ellison, M. K. Mundra, J. M. Torkelson, Impacts of Polystyrene Molecular Weight and Modification to the Repeat Unit Structure on the Glass Transition–Nanoconfinement Effect and the Cooperativity Length Scale. *Macromolecules* **38**, 1767-1778 (2005).
24. B. Jérôme, J. Commandeur, Dynamics of glasses below the glass transition. *Nature* **386**, 589-592 (1997).
25. C. J. Ellison, J. M. Torkelson, The distribution of glass-transition temperatures in nanoscopically confined glass formers. *Nature Materials* **2**, 695-700 (2003).
26. C. W. Frank *et al.*, Structure in Thin and Ultrathin Spin-Cast Polymer Films. *Science* **273**, 912-915 (1996).
27. P. A. O'Connell, G. B. McKenna, Rheological Measurements of the Thermoviscoelastic Response of Ultrathin Polymer Films. *Science* **307**, 1760-1763 (2005).
28. D. Priestley Rodney, J. Ellison Christopher, J. Broadbelt Linda, M. Torkelson John, Structural Relaxation of Polymer Glasses at Surfaces, Interfaces, and In Between. *Science* **309**, 456-459 (2005).
29. Q. Li, T. E. Tullis, D. Goldsby, R. W. Carpick, Frictional ageing from interfacial bonding and the origins of rate and state friction. *Nature* **480**, 233-236 (2011).
30. B. Thiria, M. Adda-Bedia, Relaxation Mechanisms in the Unfolding of Thin Sheets. *Physical Review Letters* **107**, 025506 (2011).
31. K. Farain, Relaxation Constant in the Folding of Thin Viscoelastic Sheets. *Physical Review Applied* **13**, 014031 (2020).
32. K. Matan, R. B. Williams, T. A. Witten, S. R. Nagel, Crumpling a Thin Sheet. *Physical Review Letters* **88**, 076101 (2002).
33. Y. Lahini, O. Gottesman, A. Amir, S. M. Rubinstein, Nonmonotonic Aging and Memory Retention in Disordered Mechanical Systems. *Physical Review Letters* **118**, 085501 (2017).
34. J. M. Hutchinson, Physical aging of polymers. *Prog. Polym. Sci.,* **20**, 703-760 (1995).

35. S. Hyun, L. Pei, J. F. Molinari, M. O. Robbins, Finite-element analysis of contact between elastic self-affine surfaces. *Physical Review E* **70**, 026117 (2004).
36. S. Dalvi *et al.*, Linking energy loss in soft adhesion to surface roughness. *Proceedings of the National Academy of Sciences* **116**, 25484 (2019).